\documentclass[preprint,prd,aps,eqsecnum,nofootinbib]{revtex4}
\usepackage{amsmath}
\usepackage{graphics}

\begin{document}

\def\cstok#1{\leavevmode\thinspace\hbox{\vrule\vtop{\vbox{\hrule\kern1pt
\hbox{\vphantom{\tt/}\thinspace{\tt#1}\thinspace}}
\kern1pt\hrule}\vrule}\thinspace}

\begin{center}
\bibliographystyle{article}
{\Large \textsc{The Eastwood--Singer gauge in Einstein spaces}}
\end{center}

\author{Giampiero Esposito$^{1}$ \thanks{
Electronic address: giampiero.esposito@na.infn.it}
Raju Roychowdhury$^{2,1}$ \thanks{
Electronic address: raju@na.infn.it}}

\affiliation{
${\ }^{1}$Istituto Nazionale di Fisica Nucleare, Sezione di Napoli,\\
Complesso Universitario di Monte S. Angelo, Via Cintia, Edificio 6, 80126
Napoli, Italy\\
${\ }^{2}$Dipartimento di Scienze Fisiche, Federico II University,\\
Complesso Universitario di Monte S. Angelo, Via Cintia, Edificio 6,
80126 Napoli, Italy}

\vspace{0.4cm}
\date{\today}

\begin{abstract}
Electrodynamics in curved space-time can be studied in the 
Eastwood--Singer gauge, which has the advantage of respecting the
invariance under conformal rescalings of the Maxwell equations. Such
a construction is here studied in Einstein spaces, for which the
Ricci tensor is proportional to the metric. The classical field 
equations for the potential are then equivalent to first solving a
scalar wave equation with cosmological constant, and then solving a
vector wave equation where the inhomogeneous term is obtained from the
gradient of the solution of the scalar wave equation. The
Eastwood--Singer condition leads to a field equation on the potential
which is preserved under gauge transformations
provided that the scalar function therein obeys a
fourth-order equation where the highest-order term is the wave operator
composed with itself. The second-order scalar equation
is here solved in de Sitter space-time, and also the fourth-order 
equation in a particular case, and these solutions are 
found to admit an exponential decay at large time provided that 
square-integrability for positive time is required. Last, the vector wave
equation in the Eastwood--Singer gauge is solved explicitly when the
potential is taken to depend only on the time variable. 
\end{abstract}

\maketitle
\bigskip
\vspace{2cm}

\section{Introduction}

Even before attempting a functional-integral quantization of gauge
theories, the analysis of hyperbolic classical field equations 
supplemented by gauge-fixing conditions is quite important. For
example, in the case of Maxwell's electrodynamics, the action functional
$-{1\over 4}\int F_{ab}F^{ab}\sqrt{-g}d^{4}x$ leads to the 
field equation
\begin{equation}
P_{a}^{\; b}A_{b}=0,
\label{(1.1)}
\end{equation}
having defined the operator
\begin{equation}
P_{a}^{\; b} \equiv -\delta_{a}^{\; b} \cstok{\ }+R_{a}^{\; b}
+\nabla_{a}\nabla^{b},
\label{(1.2)}
\end{equation}
where $\nabla$ is the Levi--Civita connection with associated Ricci
tensor $R_{a}^{\; b}$. Equation (1.1) should be supplemented by a
gauge-fixing condition \cite{1965}, i.e. an equation having general form
\begin{equation}
\Phi(A)=0,
\label{(1.3)}
\end{equation}
$\Phi$ being a functional defined on the space of gauge connection
one-forms $A_{b}dx^{b}$. In particular, 
the Lorenz-type choice \cite{PHMAA-34-287}
\begin{equation}
\Phi(A)=\Phi_{L}(A) \equiv \nabla^{b}A_{b},
\label{(1.4)}
\end{equation}
reduces Eq. (1.1) to
\begin{equation}
\left(-\delta_{a}^{\; b} \cstok{\ }+R_{a}^{\; b} \right)A_{b}=0,
\label{(1.5)}
\end{equation}
which is indeed very convenient, being the covariant vector wave equation
in curved space-time. However, the Lorenz functional has an inhomogeneous
transformation law under conformal rescalings ${\widehat g}_{ab}
=\Omega^{2} g_{ab}$ of the space-time metric, i.e. \cite{PHLTA-A107-73}
\begin{equation}
{\widehat \nabla}^{b}A_{b}=\Omega^{-2} \Bigr(\nabla^{b}A_{b}
+2 Y^{b}A_{b} \Bigr),
\label{(1.6)}
\end{equation}
having denoted by $Y_{a}$ the logarithmic derivative of the conformal
factor, i.e. $Y_{a} \equiv \nabla_{a} \log \Omega
={\nabla_{a}\Omega \over \Omega}$. There is nothing bad or undesirable
in this property, but relatively few people know that, 
in Ref. \cite{PHLTA-A107-73}, a
solution was found to the problem of achieving both field equations
and gauge-fixing condition of the conformally invariant type. For this
purpose, the idea is to add lower order curvature terms to a 
differential operator which, by itself, is not conformally invariant.
In the electromagnetic case, one therefore considers the tensor
\begin{equation}
S^{ab} \equiv -2R^{ab}+{2\over 3}R g^{ab},
\label{(1.7)}
\end{equation}
and the third-order differential operator \cite{PHLTA-A107-73}
\begin{equation}
D^{a} \equiv \nabla_{b} \Bigr( \nabla^{b}\nabla^{a}+S^{ab} \Bigr)
\label{(1.8)}
\end{equation}
acting on one-forms. From the point of view of spectral geometry and
fiber-bundle theory, the analysis of these higher-order operators is
indeed well motivated and rather natural \cite{0111003, 0309085}. 
The gauge-fixing condition imposed is
\begin{equation}
D^{b}A_{b}=0,
\label{(1.9)}
\end{equation}
and it is conformally invariant because \cite{PHLTA-A107-73}
\begin{equation}
{\widehat D}^{b}A_{b}=\Omega^{-4}\Bigr[D^{b}A_{b}
+2Y^{b}\nabla^{a}(\nabla_{a}A_{b}-\nabla_{b}A_{a})\Bigr],
\label{(1.10)}
\end{equation}
where $\nabla^{a}(\nabla_{a}A_{b}-\nabla_{b}A_{a})
=\nabla^{a}F_{ab}=0$ by virtue of the vacuum Maxwell equations imposed
upon the potential $A_{b}$. The authors of Ref. \cite{PHLTA-A107-73}
focused on background metrics which satisfy the vacuum Einstein equations.
In all dimensions greater than two, this condition is equivalent to Ricci
flatness, which in turn implies the vanishing of the tensor 
$S^{ab}$ in (1.7), while the operator in (1.8) reduces then to
$D^{a}=\cstok{\ }\nabla^{a}$, and the gauge-fixing (1.9) is implied
\cite{PHLTA-A107-73} by the Lorenz condition $\nabla^{b}A_{b}=0$.

Section 2 studies the Eastwood--Singer gauge in Einstein spaces, while its
behavior under gauge transformations of the Maxwell potential is
considered in section 3. Sections 4 and 5 solve in de Sitter space-time, 
under suitable assumptions, the associated second-order
and fourth-order scalar equations, respectively, while section 6
solves the vector wave equation in the Eastwood--Singer gauge when 
the potential depends only on the time variable. Concluding remarks
are presented in section 7. Our analysis is entirely classical, with
space-time metric of Lorentzian signature.

\section{Eastwood--Singer gauge in Einstein spaces}

The restriction to Ricci-flat space-times, however, is not mandatory, and
another interesting class of background space-times are the Einstein
spaces, for which the Ricci tensor is proportional to the metric, i.e.
$R_{ab}=\Lambda g_{ab}$. They include the de Sitter and
anti-de Sitter geometries. One then finds, in four space-time dimensions,
\begin{equation}
S^{ab}={2\over 3}\Lambda g^{ab}.
\label{(2.1)}
\end{equation}
This tensor is covariantly constant by virtue of the metric-compatible
condition $\nabla g=0$, so that the Leibniz rule and the Eastwood--Singer
gauge (1.9) yield eventually
\begin{equation}
\cstok{\ }\nabla^{b}A_{b}+S^{ab}\nabla_{b}A_{a}=0 \Rightarrow
\left(\cstok{\ }+{2\over 3}\Lambda \right)\nabla^{b}A_{b}=0.
\label{(2.2)}
\end{equation}
Moreover, the field equation (1.1) takes the form
\begin{equation}
\Bigr(-\cstok{\ }+\Lambda \Bigr)A_{a}=\nabla_{a}(-\nabla^{b}A_{b}).
\label{(2.3)}
\end{equation}
To sum up, if we define $\psi \equiv {\rm div}A=\nabla^{b}A_{b}$,
we first have to solve a scalar wave equation with cosmological
constant, i.e.
\begin{equation}
\left(\cstok{\ }+{2\over 3}\Lambda \right)\psi=0.
\label{(2.4)}
\end{equation}
The gradient of $-\psi$ provides the inhomogeneous term in the wave
equation for the Maxwell potential, i.e.
\begin{equation}
\Bigr(-\cstok{\ }+\Lambda \Bigr)A_{a}=\nabla_{a}(-\psi).
\label{(2.5)}
\end{equation}
The solutions of (2.5) can be split as
\begin{equation}
A_{a}=A_{a}^{(0)}+{\widetilde A}_{a},
\label{(2.6)}
\end{equation}
where $A_{a}^{(0)}$ obeys the homogeneous wave equation in Einstein
spaces, i.e.
\begin{equation}
\Bigr(-\cstok{\ }+\Lambda \Bigr)A_{a}^{(0)}=0,
\label{(2.7)}
\end{equation}
while ${\widetilde A}_{a}$ is a particular solution of Eq. (2.5). As a
consistency check, the sum (2.6) should fulfill Eq. (2.5). 

\section{Behavior of the Eastwood--Singer condition under gauge 
transformations}

In classical gauge theory with space-time metric of Lorentzian
signature, the gauge-fixing (also called ``supplementary'') condition
leads to a convenient form of the field equation for the potential.
For example, for classical electrodynamics in the Lorenz gauge (1.4),
the wave equation reduces to Eq. (1.5) as we said. However, while the
Maxwell Lagrangian $-{1\over 4}F_{ab}F^{ab}$ is invariant under gauge
transformations  
\begin{equation}
{ }^{f}A_{b} \equiv A_{b}+\nabla_{b}f,
\label{(3.1)}
\end{equation}
where $f$ is a freely specifiable function of class $C^{1}$, the
Lorenz gauge (1.4) (as well as any other admissible gauge) is not
invariant under (3.1). Nevertheless, to achieve the desired wave
equation on $A_{b}$, it is rather important to make sure that both
$A_{b}$ and the gauge-transformed potential
${ }^{f}A_{b}$ obey the same gauge-fixing condition, i.e.
\begin{equation}
\Phi(A)=0, \; \Phi({ }^{f}A)=0.
\label{(3.2)}
\end{equation}
A more general situation, here not considered for simplicity, is instead
the case when only the gauge-transformed potential obeys the 
gauge-fixing condition, i.e. \cite{Jackson}
\begin{equation}
\Phi(A) \not =0, \; \Phi({ }^{f}A)=0.
\label{(3.3)}
\end{equation}
From the point of view of constrained Hamiltonian systems, the 
gauge-fixing equations can be viewed as constraint equations to be
preserved following Dirac's method \cite{Hanson}. This turns a theory
with first-class constraints into a second-class set \cite{Hanson}.

The counterpart of (3.2) for pure gravity is the well known problem
of imposing the de Donder gauge on metric perturbations, and then 
requiring its invariance under infinitesimal diffeomorphisms. One
then finds \cite{08050486} that the covector 
occurring in the transformation of metric
perturbations under infinitesimal diffeomorphisms should obey a vector
wave equation (but with opposite sign of the Ricci tensor with
respect to Eq. (2.5)).

On imposing the equations (3.2), one arrives at a {\it proper subset}
of the original set of gauge transformations (3.1), where $f$ is no
longer freely specifiable, but obeys a differential equation whose
form depends on the choice of gauge-fixing functional $\Phi(A)$. In
general, $\Phi(A)$ changes under gauge transformations in a way encoded
in a differential operator of first, second or higher order.  

In our paper, by virtue of (1.9), this implies that the
function $f$ should obey the fourth-order equation
\begin{equation}
D^{b}\nabla_{b}f=0,
\label{(3.4)}
\end{equation}
which, in Einstein spaces, reads as
\begin{equation}
\left(\cstok{\ }^{2}+{2\over 3}\Lambda \cstok{\ } \right)f=0.
\label{(3.5)}
\end{equation}
Since the fourth-order operator in Eq. (3.5) is obtained from the
composition
$$
\cstok{\ } \cdot \left(\cstok{\ }+{2\over 3}\Lambda \right),
$$
the general solution can be decomposed as
\begin{equation}
f=f_{0}+f_{1},
\label{(3.6)}
\end{equation}
where $f_{0}$ belongs to the kernel of the scalar wave operator 
$\cstok{\ }+{2\over 3}\Lambda$ occurring in Eq. (2.4),
while $f_{1}$ is a particular solution of Eq. (3.5).

We might also exploit a Hodge decomposition of the potential
according to
\begin{equation}
A_{b}=A_{\perp \; b}+\nabla_{b}\phi,
\label{(3.7)}
\end{equation}
where $A_{\perp \; b}$ is the transverse part obeying
$$
\nabla^{b}A_{\perp b}=0,
$$
while $\psi \equiv \nabla^{b}A_{b}=\cstok{\ }\phi$. This would lead 
to the equations 
$$
(-\cstok{\ }+\Lambda)A_{\perp b}=0, \;
\left(\cstok{\ }+{2\over 3}\Lambda \right)\psi=0,
$$
$$
\phi=\cstok{\ }^{-1}\psi, \;
A_{b}=A_{\perp b}+\nabla_{b}(\cstok{\ }^{-1}\psi),
$$
which however are not substantially easier to solve, since our space-time
is de Sitter rather than Minkowski. We thus go on with fourth-order
equations hereafter. Note also that, if we were studying the quantum
theory via Euclidean functional integrals, our higher-order differential
operators would be elliptic operators whose one-loop contribution is
obtained from their functional determinant. We are instead studying the
gauge-fixed equations of the classical theory, e.g. Eqs. (1.5) or (2.3),
which are hyperbolic equations. Moreover, it would be misleading to 
consider the analogy between Eq. (3.4) and the equation for zero-modes
of the ghost operator of the quantum theory, since Eq. (3.4) is a
classical hyperbolic equation, and no consideration of eigenfunctions
belonging to zero-eigenvalues is involved therein.

\section{The scalar wave equation in de Sitter space-time}

The equations (2.4), (2.5) and (3.5) cannot be solved without choosing
a particular Einstein space. For this purpose, we focus on four-dimensional
de Sitter space-time in the expanding-universe coordinates, in which the
line element reads as
\begin{equation}
ds^{2}=-dt^{2}+{\rm e}^{2Ht}(dx^{2}+dy^{2}+dz^{2}),
\label{(4.1)}
\end{equation}
where the coordinates $(t,x,y,z)$ can take all values from
$-\infty$ to $+\infty$ \cite{Moller52}. This implies that
\begin{equation}
\cstok{\ }\psi=\left(-{\partial^{2}\over \partial t^{2}}
-3H{\partial \over \partial t}+{\rm e}^{-2Ht}\bigtriangleup \right)\psi,
\label{(4.2)}
\end{equation}
with $\bigtriangleup$ equal to minus the flat Laplacian in Cartesian
coordinates, i.e. (our sign convention leads to a positive-definite
leading symbol for the Laplacian)
\begin{equation}
\bigtriangleup \equiv {\partial^{2}\over \partial x^{2}}
+{\partial^{2}\over \partial y^{2}}
+{\partial^{2}\over \partial z^{2}}.
\label{(4.3)}
\end{equation}
The scalar wave equation (2.4) can be now solved by the factorization
ansatz
\begin{equation}
\psi=A(x,y,z)\chi(t),
\label{(4.4)}
\end{equation}
with the functions $A$ and $\chi$ satisfying the equations
\begin{equation}
{\rm e}^{2Ht}\left({{\ddot \chi}\over \chi}
+3H{{\dot \chi}\over \chi}-{2\over 3}\Lambda \right)
={(\bigtriangleup A)\over A}=k,
\label{(4.5)}
\end{equation}
having denoted by $k$ a positive constant. Interestingly, the second-order
ordinary differential equation for $\chi$ can be solved exactly, bearing in 
mind that $\Lambda=3H^{2}$, in the form (up to a multiplicative
parameter depending on $H$ and $k$)
\begin{eqnarray}
\chi(t)&=& {\rm e}^{-{3\over 2}Ht}\left[C_{1}\Gamma \left
(1-{\sqrt{17}\over 2} \right)
I_{-{\sqrt{17}\over 2}}\left({\sqrt{k}\over H}{\rm e}^{-Ht}\right) \right.
\nonumber \\
&+& \left . {\rm i}^{\sqrt{17}} C_{2} 
\Gamma \left(1+{\sqrt{17}\over 2}\right)
I_{{\sqrt{17}\over 2}}\left({\sqrt{k}\over H}{\rm e}^{-Ht}\right)
\right],
\label{(4.6)}
\end{eqnarray}
where $C_{1}$ and $C_{2}$ are constants and $I_{\nu}$ is the modified
Bessel function of first kind and order $\nu$. The occurrence of both
$I_{\sqrt{17}\over 2}$ and $I_{-{\sqrt{17}\over 2}}$ in (4.6) means
that the general solution contains infinitely many terms which decay
exponentially at large $t$ as well as infinitely many terms which blow
up exponentially at large $t$. The requirement of square-integrable
solutions as $t$ belongs to the positive half-line picks out 
$I_{\sqrt{17}\over 2} \left({\sqrt{k}\over H}{\rm e}^{-Ht}\right)$ in (4.6),
and enforces the choice $C_{1}=0$. Moreover, the function $A$ occurring in 
(4.4) can be expressed in the form
\begin{equation}
A(x,y,z)=\alpha(x)\beta(y)\gamma(z),
\label{(4.7)}
\end{equation}
with $\alpha,\beta,\gamma$ obeying the second-order equations
\begin{equation}
\left[{d^{2}\over dx^{2}}-\lambda_{1}\right]\alpha=0, \;
\left[{d^{2}\over dy^{2}}-\lambda_{2}\right]\beta=0, \;
\left[{d^{2}\over dz^{2}}-\lambda_{3}\right]\gamma=0,
\label{(4.8)}
\end{equation}
subject to
\begin{equation}
\lambda_{1}+\lambda_{2}+\lambda_{3}=k.
\label{(4.9)}
\end{equation}
The various choices of sign for $\lambda_{1},\lambda_{2},\lambda_{3}$
give rise to eight different combinations, which should all satisfy 
the constraint (4.10).
The solutions of Eqs. (4.9) will be oscillatory for negative 
$\lambda_{i}$ and exponentially growing for positive 
$\lambda_{i}$, i.e.
\begin{equation}
\alpha(x)=D_{1}\cos(\sqrt{|\lambda_{1}|}x) 
+D_{2}\sin(\sqrt{|\lambda_{1}|}x),
\label{(4.10)}
\end{equation}
or
\begin{equation}
\alpha(x)={\widetilde D}_{1}\cosh(\sqrt{\lambda_{1}}x) 
+{\widetilde D}_{2}\sinh(\sqrt{\lambda_{1}}x),
\label{(4.11)}
\end{equation}
and entirely analogous formulae for $\beta(y)$ and $\gamma(z)$.

\section{Solution of the fourth-order scalar equation}

As far as the fourth-order scalar wave equation (3.5) is concerned, its
form in four-dimensional de Sitter space-time with metric (4.1) reads as
(bearing again in mind that $H^{2}={\Lambda \over 3}$)
\begin{eqnarray}
\; & \; & \biggr[{\partial^{4}\over \partial t^{4}}
+6H{\partial^{3}\over \partial t^{3}}
+\left(7H^{2}-2{\rm e}^{-2Ht}\bigtriangleup \right)
{\partial^{2}\over \partial t^{2}} \nonumber \\
&-&2H({\rm e}^{-2Ht}\bigtriangleup+3H^{2}){\partial \over \partial t}
+4H^{2}{\rm e}^{-2Ht}\bigtriangleup+{\rm e}^{-4Ht}\bigtriangleup^{2}
\biggr]f=0.
\label{(5.1)}
\end{eqnarray}
With the notation of Eq. (3.4), the hard part of the analysis consists
in finding $f_{1}$, for which a factorized ansatz is not as helpful 
as for Eq. (2.4). However, we notice that Eq. (5.1) admits a particular
exact solution depending on $t$ only, because it then reduces to
\begin{equation}
\left[{d^{4}\over dt^{4}}+6H{d^{3}\over dt^{3}}
+7H^{2}{d^{2}\over dt^{2}}-6H^{3}{d\over dt}\right]f=0.
\label{(5.2)}
\end{equation}
This means we are here dealing with a proper subset of the general
set of scalar functions occurring in the gauge freedom described
by Eq. (3.1), by restricting $f$ to depend on $t$ only.
The constant-coefficient equation (5.2) is solved by the exponential
${\rm e}^{\alpha t}$, where the real parameter $\alpha$ solves the 
algebraic equation
\begin{equation}
\alpha(\alpha^{3}+6H \alpha^{2}+7 H^{2}\alpha-6H^{3})=0,
\label{(5.3)}
\end{equation}
whose four roots are
\begin{equation}
\alpha_{1}=0, \; \alpha_{2}=-3H, \;
\alpha_{3}=-{1\over 2}(3+\sqrt{17})H, \;
\alpha_{4}=-{1\over 2}(3-\sqrt{17})H.
\label{(5.4)}
\end{equation}
Equation (5.2) is therefore solved by
\begin{equation}
f(t)={\widetilde C}_{1}+{\widetilde C}_{2}{\rm e}^{-3Ht}
+{\widetilde C}_{3}{\rm e}^{-{1\over 2}(3+\sqrt{17})Ht}
+{\widetilde C}_{4}{\rm e}^{-{1\over 2}(3-\sqrt{17})Ht}.
\label{(5.5)}
\end{equation}
In this case, the requirement that $f$ should be square-integragble
for $t$ lying on the positive half-line enforces the choice 
${\widetilde C}_{1}={\widetilde C}_{4}=0$.
When we allow for a non-vanishing spatial gradient of $f$, we may expect
infinitely many terms which decay exponentially as $t \rightarrow \infty$
and are weighted by square-integrable functions of $(x,y,z)$. Their
evaluation goes beyond our present capabilities.

Indeed, Eq. (3.5) might be split into the pair of second-order equations
\begin{equation}
\left(\cstok{\ }+{2\over 3}\Lambda \right)f=v,
\label{(5.6)}
\end{equation}
\begin{equation}
\cstok{\ }v=0,
\label{(5.7)}
\end{equation}
but these are second-order equations with variable coefficients, whose
analysis is not necessarily more powerful than the single 
fourth-order equation (5.1).

\section{The vector wave equation (2.5) in de Sitter space-time}

When the line element (4.1) is exploited, the wave operator has the
following simple action on the temporal and spatial components
of the electromagnetic potential:
\begin{equation}
\cstok{\ }A_{t}=\left[-{\partial^{2}\over \partial t^{2}}
-3H{\partial \over \partial t}+e^{-2Ht} \bigtriangleup \right]A_{t},
\label{(6.1)}
\end{equation}
\begin{equation}
\cstok{\ }A_{k}=\left[-{\partial^{2}\over \partial t^{2}}
-H{\partial \over \partial t}+2H^{2}+e^{-2Ht}\bigtriangleup \right]
A_{k}-2H {\partial \over \partial x^{k}}A_{t},
\label{(6.2)}
\end{equation}
where $x^{1}=x,x^{2}=y,x^{3}=z$. We can now write down explicitly the
vector wave equation (2.5) in the Eastwood--Singer gauge, bearing in mind
that $\Lambda=3H^{2}$ therein. The associated homogeneous equations
(i.e. with vanishing right-hand side), under the assumption that
$A_{b}=A_{b}(t)$, can be solved by the ansatz (multiplicative
constants are omitted for simplicity)
\begin{equation}
A_{t}=e^{\alpha t}, \; A_{k}=e^{\beta t},
\label{(6.3)}
\end{equation}
leading to the algebraic equations
\begin{equation}
\alpha^{2}+3H \alpha +3H^{2}=0,
\label{(6.4)}
\end{equation}
\begin{equation}
\beta^{2}+H \beta+H^{2}=0.
\label{(6.5)}
\end{equation}
Hence we find
\begin{equation}
A_{t}=e^{-{3\over 2}Ht}\left[U_{1}\cos \left({\sqrt{3} \over 2}Ht \right)
+U_{2} \sin \left({\sqrt{3}\over 2}Ht \right)\right],
\label{(6.6)}
\end{equation}
\begin{equation}
A_{k}=e^{-{H\over 2}t}\left[V_{1}\cos \left({\sqrt{3} \over 2}Ht \right)
+V_{2} \sin \left({\sqrt{3}\over 2}Ht \right)\right],
\label{(6.7)}
\end{equation}
where $U_{1},U_{2},V_{1},V_{2}$ are constant coefficients.
For consistency, the function $\psi$ on the right-hand side of (2.5)
can only depend on $t$ as well, and therefore the inhomogeneous equation
for $A_{t}$ reads as
\begin{equation}
\left[{d^{2}\over dt^{2}}+3H {d\over dt}+3H^{2} \right]A_{t}
=-{d \psi \over dt},
\label{(6.8)}
\end{equation}
which is solved by
\begin{equation}
A_{t}=-\int G(t,t'){d\psi \over dt'}dt',
\label{(6.9)}
\end{equation}
having denoted by $G(t,t')$ the Green function of the second-order
operator
\begin{equation}
P_{t} \equiv {d^{2}\over dt^{2}}+3H {d\over dt}+3H^{2}.
\label{(6.10)}
\end{equation}
We therefore find $G(t,t')$ as the solution, for all $t \not = t'$,
of the equation
\begin{equation}
P_{t}G(t,t')=0,
\label{(6.11)}
\end{equation}
subject to the continuity condition
\begin{equation}
\lim_{t \to {t'}^{+}}G(t,t')=\lim_{t \to {t'}^{-}}G(t,t'),
\label{(6.12)}
\end{equation}
jointly with the jump condition
\begin{equation}
\lim_{t \to {t'}^{+}} {\partial G \over \partial t}
-\lim_{t \to {t'}^{-}} {\partial G \over \partial t}=1,
\label{(6.13)}
\end{equation}
and the `boundary conditions'
\begin{equation}
B_{1}G \equiv \alpha_{11}G(t_{0},t')+\alpha_{12}{\dot G}(t_{0},t')=0,
\label{(6.14)}
\end{equation}
\begin{equation}
B_{2}G \equiv \alpha_{21}G(t_{1},t')+\alpha_{22}{\dot G}
(t_{1},t')=0.
\label{(6.15)}
\end{equation}
If we denote by $s_{1}(t)$ and $s_{2}(t)$ the solutions of the problems
\begin{equation}
s_{1}=u: \; P_{t}u(t)=0, \; B_{1}u(t_{0})=0,
\label{(6.16)}
\end{equation}
\begin{equation}
s_{2}=u: \; P_{t}u(t)=0, \; B_{2}u(t_{1})=0,
\label{(6.17)}
\end{equation}
we can write
\begin{equation}
G(t,t')=A(t')s_{1}(t) \; \forall t \in ]t_{0},t' [,
\label{(6.18)}
\end{equation}
\begin{equation}
G(t,t')=B(t')s_{2}(t) \; \forall t \in ]t',t_{1}[ .
\label{(6.19)}
\end{equation}
The conditions (6.12) and (6.13) lead therefore to \cite{Stakgold}
\begin{equation}
G(t,t')={s_{1}(t_{<})s_{2}(t_{>}) \over W(s_{1},s_{2};t')},
\label{(6.20)}
\end{equation}
where $t_{<} \equiv {\rm min}(t,t'), t_{>} \equiv {\rm max}(t,t')$,
and $W$ is the Wronskian
\begin{equation}
W(s_{1},s_{2};t')=s_{1}(t'){\dot s}_{2}(t')
-{\dot s}_{1}(t')s_{2}(t').
\label{(6.21)}
\end{equation}
Now Abel's formula for the Wronskian tells us that \cite{Stakgold}
\begin{equation}
W(s_{1},s_{2};t')={\rm cont.} \times e^{-s(t')},
\label{(6.22)}
\end{equation}
where, bearing in mind (6.10), $s$ is a particular solution of
the equation
\begin{equation}
{ds \over dt'}=3H,
\label{(6.23)}
\end{equation}
i.e. $s(t')=3Ht'+{\rm constant}$. On denoting by $\kappa$ an
integration constant, we arrive at the desired formula
\begin{equation}
G(t,t')=\kappa e^{3Ht'} s_{1}(t_{<}) s_{2}(t_{>}).
\label{(6.24)}
\end{equation}
For example, on choosing the boundary conditions (cf. (6.14)--(6.17))
\begin{equation}
s_{1}(t_{0})=0, \; s_{2}(t_{1})=0,
\label{(6.25)}
\end{equation}
we obtain
\begin{equation}
s_{1}(t)=e^{-{3\over 2}Ht} \left[
\cos \left({\sqrt{3}\over 2}Ht \right)
-\cot \left({\sqrt{3}\over 2}H t_{0}\right)
\sin \left({\sqrt{3}\over 2}Ht \right)\right],
\label{(6.26)}
\end{equation}
\begin{equation}
s_{2}(t)=e^{-{3\over 2}Ht} \left[
\cos \left({\sqrt{3}\over 2}Ht \right)
-\cot \left({\sqrt{3}\over 2}H t_{1}\right)
\sin \left({\sqrt{3}\over 2}Ht \right)\right],
\label{(6.27)}
\end{equation}
and we have only to plug (6.26) and (6.27) into (6.24) to obtain the 
explicit form of the Green function occurring in (6.9).     
Moreover, $\psi(t)$ in (6.8) is the combination of modified Bessel
functions of Eq. (4.6), by virtue of (2.4).

The desired solution of Eq. (2.5) for $A_{t}$ is therefore given by
the sum of (6.6), which solves the homogeneous equation, and (6.9),
the latter being a solution of the inhomogeneous equation (6.8).
Interestingly, thanks to the cosmological constant of de Sitter
space-time, all time-dependent solutions of our vector wave equation
(2.5) are square-integrable for $t \in (0, \infty)$.

When the potential $A_{b}$ has non-vanishing spatial gradient,
our particular solutions remain helpful, because one can then look
for exact or asymptotic solutions where the constant coefficients,
e.g. $U_{1},U_{2},V_{1},V_{2}$ in (6.6) and (6.7), are promoted to
position-dependent functions \cite{Rendall}.

\section{Concluding remarks}

The Eastwood--Singer gauge \cite{PHLTA-A107-73}, whose physical
motivations have been described in Section 1, has its deep mathematical
roots in the theory of conformally covariant operators 
\cite{0111003, 0309085}. However, explicit calculations in curved
geometries of cosmological interest were lacking, to the best of our
knowledge. In our paper we have first reduced its analysis in Einstein
spaces to studying Eqs. (2.4), (2.5), (3.5), and then we have found exact
solutions of the scalar wave equation (2.4) and a particular
solution of the scalar equation (3.5). Moreover, the solution of
the inhomogeneous vector wave equation (2.5) has been obtained when
the potential depends only on the time variable. It now remains to be
seen when the technique in Ref. \cite{Rendall} or, instead, in Ref.
\cite{PHRVA-D10-1070}, can be used to solve Eq. (2.5) when the potential
$A_{b}$ depends on all space-time coordinates.

All of this leaves aside quantization issues, for which we refer the
reader to the work in Ref. \cite{PHRVA-D56-2442}.

\acknowledgments

The authors are grateful to the Dipartimento di Scienze Fisiche
of Federico II University, Naples, for hospitality and support.
We are indebted to Ebrahim Karimi for computer assistance.


\begin{references}
\bibitem{1965}
B. S. DeWitt, {\it Dynamical Theory of Groups and Fields}
(Gordon \& Breach, New York, 1965).
\bibitem{PHMAA-34-287}
L. Lorenz, {\it Phil. Mag.} {\bf 34} (1867) 287. 
\bibitem{PHLTA-A107-73}
M. Eastwood and M. Singer, {\it Phys. Lett. A} {\bf 107} (1985) 73. 
\bibitem{0111003}
T. Branson and A. R. Gover, arXiv:hep-th/0111003.
\bibitem{0309085}
T. Branson and A. R. Gover, arXiv:math/0309085.
\bibitem{Jackson}
J. D. Jackson, {\it Classical Electrodynamics} (Wiley, New York, 1975).
\bibitem{Hanson}
A. Hanson, T. Regge and C. Teitelboim, {\it Constrained Hamiltonian
Systems}, Contributi del Centro Linceo Interdisciplinare di Scienze
Matematiche e loro Applicazioni, Vol. {\bf 22} (Accademia Nazionale
dei Lincei, Rome, 1976).
\bibitem{08050486}
D. Bini, S. Capozziello and G. Esposito, arXiv:0805.0486 (to appear in
{\it Int. J. Geom. Meth. Mod. Phys.}).
\bibitem{Moller52}
C. Moller, {\it The Theory of Relativity} (Clarendon Press, Oxford, 1952).
\bibitem{Stakgold}
I. Stakgold, {\it Green's Functions and Boundary Value Problems}
(Wiley, New York, 1979).
\bibitem{Rendall}
A. Rendall, {\it Ann. Henri Poincar\'e} {\bf 5} (2004) 1041.
\bibitem{PHRVA-D10-1070}
J. M. Cohen and L. S. Kegeles, {\it Phys. Rev. D} {\bf 10} (1974) 1070. 
\bibitem{PHRVA-D56-2442}
G. Esposito, {\it Phys. Rev. D} {\bf 56} (1997) 2442. 
\end{references}
\end{document}